\documentclass[aps, prd,twocolumn,superscriptaddress, groupedaddress]{revtex4}
\usepackage{graphicx}  
\usepackage{dcolumn}   
\usepackage{bm}        
\usepackage{amssymb}   
\usepackage{amsmath}
\usepackage{hyperref}
\usepackage{cleveref}
\usepackage{subcaption}
\usepackage{soul}

\setstcolor{red} 

\crefformat{figure}{Fig.~#2#1#3}
\crefformat{equation}{equation~(#2#1#3)}
\crefformat{section}{section~#2#1#3}
\crefformat{table}{Tab.~#2#1#3}
\crefformat{appendix}{appendix~#2#1#3}
\crefmultiformat{figure}{Figs.~#2#1#3}{~and~#2#1#3}%
    {,~#2#1#3}{,~#2#1#3}
\crefrangeformat{figure}{Figs.~(#3#1#4--#5#2#6)}

\newcommand{\comment}[1]{}

\usepackage[T1]{fontenc}
\usepackage{ae,aecompl}
\usepackage{newtxtext,newtxmath}

\usepackage[colorinlistoftodos]{todonotes}

\begin{document}
\title{Calibrating Bayesian Tension Statistics using Neural Ratio Estimators}

\author{Harry T. J. Bevins}
\email{htjb2@cam.ac.uk}
\affiliation{Astrophysics Group, Cavendish Laboratory, Cambridge, CB3 0HE, UK}
\affiliation{Kavli Institute for Cosmology, Cambridge, CB3 0HA, UK}

\author{William J. Handley}
\affiliation{Kavli Institute for Cosmology, Cambridge, CB3 0HA, UK}
\affiliation{Institute of Astronomy, Cambridge, CB3 0HA, UK}

\author{Thomas Gessey-Jones}
\affiliation{Astrophysics Group, Cavendish Laboratory, Cambridge, CB3 0HE, UK}
\affiliation{Kavli Institute for Cosmology, Cambridge, CB3 0HA, UK}

\begin{abstract}

When fits of the same physical model to two different datasets disagree, we call this tension. Several apparent tensions in cosmology have occupied researchers in recent years, and a number of different metrics have been proposed to quantify tension. Many of these metrics suffer from limiting assumptions, and correctly calibrating these is essential if we want to successfully determine whether discrepancies are significant.
A commonly used metric of tension is the evidence ratio $R$. The statistic has been widely adopted by the community as a Bayesian way of quantifying tensions, however, it has a non-trivial dependence on the prior that is not always accounted for properly.
We show that this can be calibrated out effectively with Neural Ratio Estimation.
We demonstrate our proposed calibration technique with an analytic example, a toy example inspired by 21-cm cosmology, and with observations of the Baryon Acoustic Oscillations from the Dark Energy Spectroscopic Instrument~(DESI) and the Sloan Digital Sky Survey~(SDSS). We find no significant tension between DESI and SDSS. 
\end{abstract}

\maketitle

\section{Introduction}

Independently confirming conclusions about the nature of our Universe from one experiment with another is crucial to the advance of knowledge. When the inference from two different experiments disagree with each other, we call this tension.
Tension between different datasets raises questions about the need for new physics and better descriptions of our instruments and systematics. The $H_0$ and $\sigma_8$ tensions \citep[e.g.][]{knox2020HubbleHunters, Efstathiou2020LockdownH0, Amon2022s8, Preston2023s8, Abbott2023DESKIDS}  are the most commonly encountered examples in cosmology, but other examples include observations of the 21-cm signal from cosmic dawn \citep{Singh_SARAS3_2022}, tensions in the amplitude of the matter power spectrum \citep{Battye_tensions_2015} and tension in estimates of the curvature of the Universe \cite{Handley2019Curvature}. A historical example of tension is in the measurement of the matter density $\Omega_m$ \cite{Peebles1984OmegaM, Krauss1995OmegaM, Ostriker1995OmegaM, Efstathiou1990OmegaM} which was resolved by the discovery of the accelerating universe \cite{Riess1998Accelerating}. A detailed review of cosmological tensions can be found in \cite{Abdalla2022TensionReview}. 

Many different measures of tension have been proposed and are widely used in cosmological studies. Examples include Bayesian Suspiciousness \cite{Handley_tensions_2019}, estimators of the probability of observed parameter differences \cite{Raveri2020parameterDifference, Raveri2021ParameterDiff}, Goodness of fit degradation \cite{Raveri2019GoFDegredation} and Eigentension \cite{Park2020Eigentension}. These tension statistics are summarised and reviewed in \cite{Charnock_2017, Lemos2021DESTensions} and \cite{Saraivanov2024Metrics}. It is common practice to rephrase these tension metrics in $\sigma$ units of tension,  corresponding to probabilities on a one dimensional normal distribution. When expressed in this way, one would expect that the various tension metrics predict the same level of tension or concordance between different datasets. However, this is often not the case due to the various assumptions that are made when defining the statistics.
To tackle this issue, we can try to define tension metrics that do not make these assumptions, however, the tension statistics often lose some of their interpretability when we do this. Instead, we try to calibrate out these assumptions in sensible ways. Calibration of tension statistics is an important and often overlooked step that needs to be taken to correctly interrogate the tension between different datasets.

A commonly used metric of tension is $R$ corresponding to the ratio of a joint evidence and the product of individual evidences for two different datasets under a common model. The $R$ statistic was first proposed in \citep{Marshall_2006} and has been used to quantify tension in a number of cosmological studies \citep[e.g.][]{Trotta_2008, Seehars_2016}. $R$ has also been used to perform model comparison in some works, although the authors of \cite{Cortes2024Tension} showed that this approach to model comparison is incomplete. 

The ratio $R$ suffers from a non-trivial dependence on the prior that is not always accounted for properly \citep{Handley_tensions_2019}. In \cite{Handley_tensions_2019} the authors showed that as one decreases the prior width on the common parameters in the model of the two datasets, then the tension between the datasets should increase and $R$ should decrease. Intuitively, one can see that it is more satisfying if the two experiments favour parameters that are close together given a wide prior in comparison to a narrow prior. However, what constitutes `narrow' and `wide' is problem specific and subjective, making the interpretation of $R$ difficult. We would like to calibrate out the prior dependence.

The authors of \cite{Handley_tensions_2019} propose an alternative statistic that is closely related to $R$ called the Suspiciousness $S$. Suspiciousness is calculated by subtracting a correction from $R$ that quantifies the probability that the data sets would match given the prior ranges on the model parameters. It can be thought of as the value of $R$ in the case where the priors are as narrow as possible without impacting the shape of the posterior. In \cite{Cuceu2019BAO} the authors convert $S$ into $\sigma$s of tension, however, this requires an estimate of the number of constrained dimensions $d$ in the joint analysis. To estimate $d$ they use the Bayesian (sometimes referred to as Gaussian) Model Dimensionality, however, this is a poor estimator of $d$ if the posterior is significantly non-Gaussian, as is often the case in cosmology.

In \cite{Lemos2021DESTensions} the authors demonstrate that simulations can be used to calibrate tension metrics with the Planck data and Dark Energy Survey~(DES) data. They proposed taking a fiducial set of parameters, such as the maximum posterior point for Planck, shifting these parameter values by some posterior-informed step sizes to induce a known degree of tension, simulate the now in tension DES observations and calculate the value of ones chosen tension metric between the real Planck data and the simulation. 
To do this, however, one often has to run expensive sampling algorithms on the simulated data to calculate tension statistics such as $R$ and $S$.

We propose calibrating the prior dependence of $R$ using neural ratio estimation~(NRE) \citep[e.g.][]{Cranmer202SBI, Miller2021TMNRE, Cole2022TMNRE}.  
NREs are classifiers, with interpretable outputs, that determine whether two quantities are drawn from independent distributions or a joint distribution. 
We show that the output of an NRE trained on simulations of two experiments observables can be interpreted as $R$ and that an appropriately trained NRE can be used to calibrate for the prior dependence of the dimensionless $R$ statistic. Since $R$ requires the calculation of three Bayesian evidences, it is an expensive statistic to evaluate using traditional methods like nested sampling. We show that $R$ can be accessed with a significantly smaller computational overhead using cutting edge machine learning tools.
We call our NRE setup the \textsc{tensionnet}.

In \cref{sec:bayes} we summarise Bayesian inference and give more details about $R$. In \cref{sec:interpretation} we discuss the interpretation of $R$ and follow this with a discussion on NREs in \cref{sec:nres}. We discuss calibrating $R$ with NREs in \cref{sec:calibration_r}. We then test our method on toy examples with known in concordance $R$ distributions in \cref{sec:validation}. We then apply the \textsc{tensionnet} to a toy example inspired by 21-cm cosmology and to assess the tension between Baryon Acoustic Oscillations~(BAO) observations from the Dark Energy Spectroscopic Instrument~(DESI) and the Sloan Digital Sky Survey~(SDSS), in \cref{sec:cosmological_examples}. We consider some limitations of our method in \cref{sec:limitations} and conclude in \cref{sec:conclusions}.

The code used in this paper is publicly available at \url{https://github.com/htjb/tension-networks}.

\section{Bayesian Inference and Tension Statistics}
\label{sec:bayes}

In Bayesian inference we are interested in modelling data $D$ with a model $M$ containing parameters $\theta$ to recover both the probability of the data given the model $\mathcal{Z} = P(D|M)$ or evidence and the probability of a given $\theta$ given the data and model $P(\theta) = P(\theta|D, M)$ or posterior. To do this we draw samples from a prior $\pi(\theta)=P(\theta|M)$ which encodes our prior knowledge of the parameters and evaluate a likelihood which is our postulated probability of the data given a set of parameters and model $L(\theta) = P(D|\theta, M)$. We relate these quantities using Bayes theorem
\begin{equation}
    P(\theta|D, M) = \frac{P(D|\theta, M)P(\theta|M)}{P(D|M)} = \frac{\mathcal{L}(\theta) \pi(\theta)}{\mathcal{Z}},
    \label{eq:bayes}
\end{equation}
where $\mathcal{Z}$ is given by
\begin{equation}
    \mathcal{Z} = \int \mathcal{L}(\theta) \pi(\theta) d\theta.
\end{equation}
An efficient and accurate way to recover both the evidence and the posterior is with nested sampling \citep{skilling_ns_2006, Ashton_NS_2022} although other methods exist \citep[e.g.][]{Trotta_2007, Heavens_2017, Srinivasan2024floZ, emcee, Polanska2023harmonicmean}.

The tension $R$ between two datasets, indicated by the subscripts $A$ and $B$, is
\begin{equation}
    R = \frac{\mathcal{Z}_{A,B}}{\mathcal{Z}_A \mathcal{Z}_B} = \frac{P(D_A, D_B|M)}{P(D_A|M)P(D_B|M)} = \frac{P(D_A, D_B)}{P(D_A)P(D_B)},
    \label{eq:R_statistic}
\end{equation}
where we have dropped the dependence on $M$ in the last expression for conciseness.
$R$ is prior dependent and this can be seen by noting that
\begin{equation}
\begin{aligned}
    R &= \frac{\mathcal{Z}_{A,B}}{\mathcal{Z}_A \mathcal{Z}_B} = \frac{1}{\mathcal{Z}_A \mathcal{Z}_B} \int \mathcal{L}_A \mathcal{L}_B \pi d\theta \\ & = \int \frac{\mathcal{L}_A \pi}{\mathcal{Z}_A} \frac{\mathcal{L}_B \pi}{\mathcal{Z}_B} \frac{1}{\pi} d\theta  = \int \frac{P_A P_B}{\pi} d\theta \\ & = \bigg\langle \frac{P_B}{\pi} \bigg\rangle_{P_A} =  \bigg\langle \frac{P_A}{\pi} \bigg\rangle_{P_B}
\end{aligned}
    \label{eq:prior_dependence}
\end{equation} 
where we have assumed the data sets are independent, and the angled brackets represent averages over the distributions $P_A$ and $P_B$ \citep{Handley_tensions_2019}. For a uniform prior, $\pi = 1/V$ where $V$ is the volume, one can see from \cref{eq:prior_dependence} that if the prior is made smaller than $R$ being proportional to $V$ also decreases. This logic generalises to more complicated priors.

\section{Interpreting $R$}
\label{sec:interpretation}

$R$ has several attractive properties \cite{Handley_tensions_2019};
\begin{itemize}
    \item \textbf{Dimensionally consistent:} Whilst the Bayesian evidence typically carries units that are the inverse of the measure on the data space, by taking the ratio of evidences the units cancel. $R$ is therefore not dependent on the units of the model space.
    \item \textbf{Parameterisation Invariant:} The value of $R$ does not change if you reparameterise the model (e.g. switch from mass to log mass). While priors and likelihoods can change under reparameterisations the Bayesian evidence should properly account for this and ratios of evidences will remain invariant to these changes. $R$ therefore reflects the true model-data compatibility and is not sensitive to arbitrary coordinate choices.
    \item \textbf{Symmetric:} Swapping the two data sets does not affect the value of $R$. So there is no preferred ordering and both data sets are treated equally.
\end{itemize}
It is typically interpreted with respect to a value of 1 with $R \ll 1$ corresponding to inconsistent datasets and $R \gg 1$ to consistent data. However, this interpretation does not tell us the degree to which our datasets are in tension given the prior and model choice. To try and quantify the tension between observations from the Dark Energy Survey and Planck, the authors of \cite{DES_Y1} interpreted $R$ on a Jefferys' scale\citep{jeffreys1983theory}. The Jefferys' scale is, however, somewhat arbitrary.

In \cite{Handley_tensions_2019} the authors showed that
\begin{equation}
    R = \frac{\mathcal{Z}_{A,B}}{\mathcal{Z}_A \mathcal{Z}_B} = \frac{P(D_A, D_B)}{P(D_A)P(D_B)} =\frac{P(D_A|D_B)}{P(D_A)} = \frac{P(D_B|D_A)}{P(D_B)},
\end{equation}
implying then one can interpret $R$, if it is greater than $1$, as a fractional increase in confidence in dataset $A$ given knowledge of dataset $B$ over $A$ alone (or vice versa). If $R \ll 1$ then the authors suggest we should be concerned about our model or the datasets.

When interpreting $R$ one has to keep in mind the impact which the prior has on its value. Reducing the width of the prior will increase the apparent tension between the datasets by reducing the value of $R$. The authors of \cite{Handley_tensions_2019} suggest repeating our analysis with sensible modifications to the prior distribution to determine how stable the value of $R$ is and consequently our conclusions regarding the tension between different datasets.

Given a choice of prior and model, there is a distribution of possible in concordance $R$ values that could be observed between two different experiments. Low signal-to-noise observations of the same signal by the two experiments will have lower typical values of $R$ in contrast to high signal-to-noise observations. This distribution can be used to translate between $R$ and $N\sigma$ estimates of tension, removing the prior dependence from the statistic and allowing for comparison with other tension metrics. The difficulty, however, is in accessing this distribution, which requires the evaluation of individual and joint evidences for a large sample of simulations, making it computationally expense and often intractable. In this paper, we propose calibrating the prior dependence of $R$ using simulations and NREs to quickly evaluate the in concordance $R$ distribution.

\section{Neural Ratio Estimation}
\label{sec:nres}

\begin{figure*}
    \centering
    \begin{tikzpicture}[rednode/.style={circle, draw=red!60, fill=red!5, very thick, minimum size=5mm},
                bluenode/.style={circle, draw=blue!60, fill=blue!5, very thick, minimum size=5mm},
                greennode/.style={circle, draw=green!60, fill=green!5, very thick, minimum size=5mm},
                node distance=0.2cm and 1.7cm,
                remember picture]
        \node (0) {};
        
        \node[rednode, above=of 0, text width=0.5cm, align=center](layer1_center1) {};
        \node[rednode, below=of layer1_center1, text width=0.5cm, align=center](layer1_center2) {};
        \node[rednode, above=of layer1_center1, text width=0.5cm, align=center](layer1_top2) {};
        \node[rednode, below=of layer1_center2, text width=0.5cm, align=center](layer1_bottom1) {};
    
        \node[rednode, left=of layer1_top2, text width=0.5cm, align=center](hl1) {};
        \node[rednode, left=of layer1_center1, text width=0.5cm, align=center](hl2) {};
        \node[rednode, left=of layer1_center2, text width=0.5cm, align=center](hl3) {};
        \node[rednode, left=of layer1_bottom1, text width=0.5cm, align=center](hl4) {};
        
        \node[bluenode, below left=of hl1, text width=0.5cm, align=center, yshift=0.5cm](input_1) {$D_A$};
        \node[above=of input_1]{\Huge $\vdots$};
        \node[below=of input_1]{\Huge $\vdots$};
        \node[bluenode, below=of input_1, text width=0.5cm, align=center, yshift=-1cm](input_2) {$D_B$};
        \node[above=of input_2]{\Huge $\vdots$};
        \node[below=of input_2]{\Huge $\vdots$};

        \node[greennode, right=of layer1_center1, text width=0.7cm, align=center, yshift=-0.65cm](output_1) {$\log R$};
        \node[draw, rectangle, right=of output_1, text width=2cm, align=center, xshift=-1cm](output_2) {$p = S_\sigma(\log R)$};

        \node[draw, rectangle, below=of output_1, text width=7cm, align=center,
        xshift=2cm](loss) {
        Loss Function:
        \begin{equation*}
        l = \frac{1}{N}\bigg[ \sum_{i}^N y_i \log (p_i) + (1-y_i)\log (1 - p_i)\bigg]
        \end{equation*}};

        \node[draw, rectangle, above=of layer1_top2, text width=7cm, align=center, xshift=1cm](title){
        Neural Ratio Estimation of $\log R = \log \frac{P(D_A, D_B)}{P(D_A)P(D_B)}$
        };

        \draw[->, dashed](output_1.east) -- (output_2.west);
        \draw[->](input_1.east) -- (hl1.west);
        \draw[->](input_1.east) -- (hl2.west);
        \draw[->](input_1.east) -- (hl3.west);
        \draw[->](input_1.east) -- (hl4.west);
        
        \draw[->](input_2.east) -- (hl1.west);
        \draw[->](input_2.east) -- (hl2.west);
        \draw[->](input_2.east) -- (hl3.west);
        \draw[->](input_2.east) -- (hl4.west);

        \draw[->](hl1.east) -- (layer1_center1.west);
        \draw[->](hl2.east) -- (layer1_center1.west);
        \draw[->](hl3.east) -- (layer1_center1.west);
        \draw[->](hl4.east) -- (layer1_center1.west);
        
        \draw[->](hl1.east) -- (layer1_center2.west);
        \draw[->](hl2.east) -- (layer1_center2.west);
        \draw[->](hl3.east) -- (layer1_center2.west);
        \draw[->](hl4.east) -- (layer1_center2.west);
        
        \draw[->](hl1.east) -- (layer1_top2.west);
        \draw[->](hl2.east) -- (layer1_top2.west);
        \draw[->](hl3.east) -- (layer1_top2.west);
        \draw[->](hl4.east) -- (layer1_top2.west);
        
        \draw[->](hl1.east) -- (layer1_bottom1.west);
        \draw[->](hl2.east) -- (layer1_bottom1.west);
        \draw[->](hl3.east) -- (layer1_bottom1.west);
        \draw[->](hl4.east) -- (layer1_bottom1.west);
        
        \draw[->](layer1_center1.east) -- (output_1.west);
        \draw[->](layer1_center2.east) -- (output_1.west);
        \draw[->](layer1_top2.east) -- (output_1.west);
        \draw[->](layer1_bottom1) -- (output_1.west);

    \end{tikzpicture}
    \caption{A schematic of the neural ratio estimator~(NRE) used in this work, which we refer to as a \textsc{tensionnet}. The NRE is trained on matched and mismatched pairs of simulated observations from two different experiments $A$ and $B$ and outputs an estimate of the tension statistic $R$. The network is trained using the binary cross entropy loss function.}
    \label{fig:tension_network}
\end{figure*}
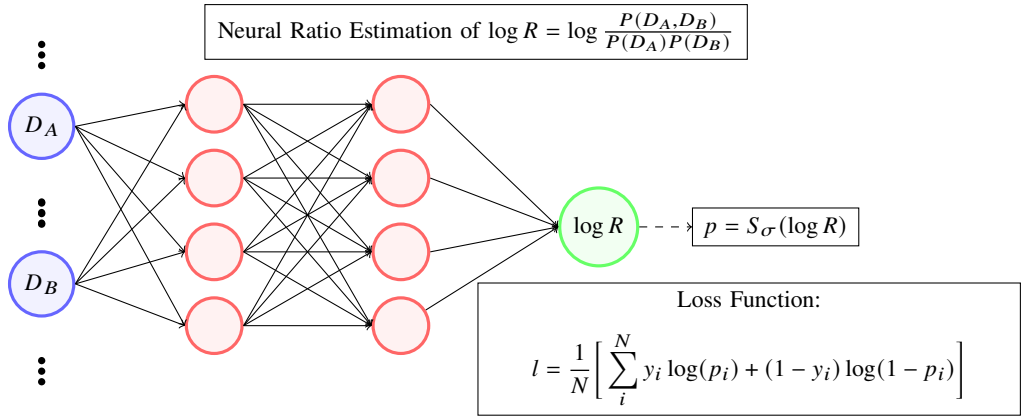

Neural Ratio Estimators~(NRE) are neural network classifiers that  are trained to return the probability that two inputs have been drawn from a joint distribution relative to the probability that they have been drawn from independent distributions. For training data that includes an equal number of examples of two inputs $A$ and $B$ drawn from their independent distributions and their joint distribution, the output of a neural ratio estimator tends towards
\begin{equation}
    \log r = \log \frac{P(A, B)}{P(A)P(B)}.
\end{equation}

To prove this, we begin by defining the network output as $f(A, B)$. During training, we give it examples drawn from the joint distribution $P(A, B)$ with probability $P_{\rm J}$ and drawn from $P(A)P(B)$ with probability $(1 - P_{\rm J})$. NREs are trained with a binary cross entropy loss function that is defined as
\begin{equation}
     l = \frac{1}{N}\bigg[ \sum_{i}^N y_i \log (\tilde{f}(A, B)) + (1-y_i)\log (1 - \tilde{f}(A, B))\bigg],
\end{equation}
where 
\begin{equation}
    \tilde{f}(A, B) \equiv S_\sigma(f(A, B)) = \frac{e^{f(A, B)}}{1 + e^{f(A, B)}},
\end{equation}
and where $y_{\rm i}$ is 1 for samples drawn from the joint and 0 for independent samples. $S_\sigma$ is the sigmoid activation function and scales the output of the network between 0 and 1.

In the limit of a large number of training samples, we can take the continuous limit of the sum
\begin{equation}
\begin{aligned}
    l \approx &- \int P(A, B) P_{\rm J} \log(\tilde{f}(A, B)) \\ &
    + P(A) P( B) (1 - P_{\rm J}) \log(1 - \tilde{f}(A, B)) dA dB.
\end{aligned}
\end{equation}
where the approximation approaches equality as the size of the training data set approaches infinity. During training the loss function is minimized and so we can find the function the network should converge to via the calculus of variations
\begin{equation}
    0 = \frac{\delta l}{\delta \tilde{f}} = \frac{P(A, B) P_{\rm J} }{\tilde{f}(A, B)}  - \frac{P(A) P( B) (1 - P_{\rm J})}{1 - \tilde{f}(A, B)},
\end{equation}
which can be rewritten as
\begin{equation}
    \tilde{f}(A, B) = \frac{\frac{P(A, B)P_{\rm J}}{ P(A) P( B) (1 - P_{\rm J})}}{1 + \frac{P(A, B)P_{\rm J}}{ P(A) P( B) (1 - P_{\rm J})}}.
\end{equation}
Recalling that the output of our network is defined such that $\tilde{f}(A, B) = S_\sigma(f(A, B))$ we see that 
\begin{equation}
    f(A, B) \rightarrow \log\left(\frac{P(A, B)P_{\rm J}}{ P(A) P( B) (1 - P_{\rm J})} \right),
\end{equation}
which when $P_{\rm J} = 0.5$ gives
\begin{equation}
    f(A, B) \rightarrow \log r,
\end{equation}
where in the limit of perfect training $f(A, B) = \log r $.

\section{Calibrating $R$ with NREs}
\label{sec:calibration_r}

As discussed above, a trained NRE outputs the log of the ratio
\begin{equation}
    r = \frac{P(A,B)}{P(A)P(B)}.
\end{equation}
It can be seen, trivially,
\begin{equation}
    r = R = \frac{P(D_A,D_B)}{P(D_A)P(D_B)} = \frac{\mathcal{Z}_{A,B}}{\mathcal{Z}_A \mathcal{Z}_B},
\end{equation}
if the inputs to the NRE $A$ and $B$ correspond to the datasets $D_A$ and $D_B$

We propose that the true observed tension $R_\mathrm{obs}$ is calculated using nested sampling \citep[e.g.][]{Handley_tensions_2019} or an alternative independent evidence estimation tool. Then we propose using the NRE to predict the in concordance $R$ distribution, against which one can calibrate $R_{\rm obs}$. A schematic of the NRE or \textsc{tensionnet} is shown in \cref{fig:tension_network}.

\begin{figure*}
    \centering
    \includegraphics{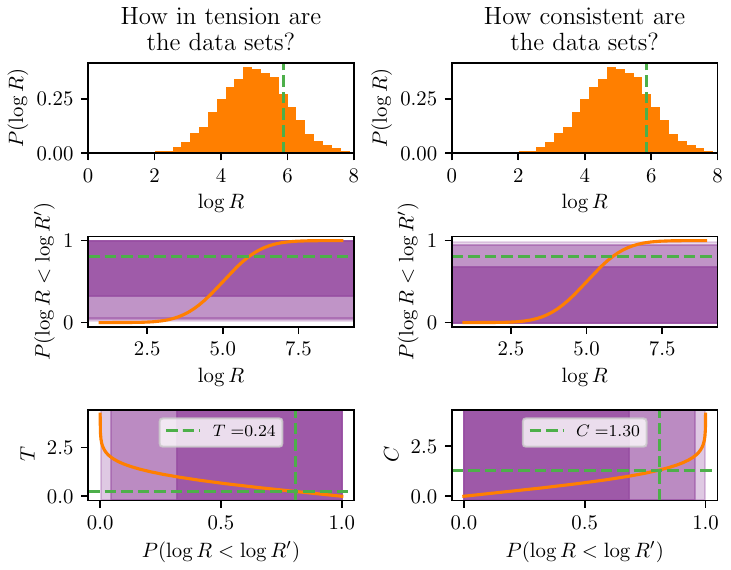}
    \caption{Interpreting $R_\mathrm{obs}$ with NREs. The top row of the figure shows an example distribution of possible in concordance $R$ values. As we move to the right of the median of the distribution we move towards concordance and to the left, lower values of $\log R$, towards tension.
    The middle row of the figure shows the corresponding cumulative distribution function, and the bottom row shows how the tension statistic $T$ and concordance statistic $C$ vary with $\log R$ for this example. The observed $\log R_\mathrm{obs}$, its corresponding value on the CDF and its value on $T$ and $C$ are shown as green dashed lines. The shaded regions show the 1,2 and 3 $\sigma$ contours for both statistics with the darker region representing 1$\sigma$ and the lighter region 3$\sigma$.}
    \label{fig:interpretation}
\end{figure*}

In practice, our proposed calibration method is as follows;
\begin{enumerate}
    \item Generate a set of matched simulations, using the same models and prior used to evaluate $R_{\rm obs}$, of $D_A(\theta)$ and $D_B(\theta)$ where they share the same parameters. This gives us the set $s =\{D_A(\theta_i), D_B(\theta_i)\}_{i=0}^N$.
    \item We then shuffle one set of the simulations to give us $s^\prime = \{D_A(\theta_i), D_B(\theta_j)\}_{i \neq j = 0}^{N}$.
    \item We label the matched sets of data with a value of 1 and the mismatched data with a value of 0.
    \item We then shuffle our labelled matched and mismatched data and split this into training and validation data.
    \item We then train our Neural Ratio Estimator and perform early stopping using the validation data.
    \item Once trained we then generate a new set of matched datasets, $z =\{D_A(\theta_i), D_B(\theta_i)\}_{i=0}^N$, from the models covering the entire prior range and calculate their corresponding $\log R$ values with the NRE to recover the in concordance distribution.
    \item Given samples on this distribution $P(\log R)$ we then calculate an empirical CDF, $P(\log R < \log R^\prime)$ which along with the inverse survival function of the standard normal distribution can be used to translate $R$ into the desired prior calibrated $N\sigma$ measure of tension.
\end{enumerate}

The inverse survival function $z(\alpha)$ is defined as the probability that a random variable $X$ takes a value less than $x$.  Specifically, we are interested in the one-sided inverse survival function which for a standard normal distribution is
\begin{equation}
    z\bigg(\frac{\alpha}{2}\bigg) = \sqrt{2}\mathrm{erf}^{-1}\bigg(2 \bigg(1 - \frac{\alpha}{2}\bigg) -1\bigg).
\end{equation}

We can define a prior calibrated tension statistic from the CDF of the $\log R$ distribution
\begin{equation}
\begin{aligned}
    T = z\bigg(\frac{P(\log R < \log R^\prime)}{2}\bigg) = \sqrt{2}\mathrm{erf}^{-1}(1 - P(\log R < \log R^\prime)),
\end{aligned}
\end{equation}
If $P(\log  R< \log R^\prime) = 1$ then $T = 0$ and we should be concerned that are datasets are in perfect agreement. If $T=3$ for example, then we can say that the experiments are in $3\sigma$ tension. Conversely, we can define a concordance statistic
\begin{equation}
\begin{aligned}
    C & = z\bigg(\frac{(1-P(\log R < \log R^\prime))}{2}\bigg) \\& = \sqrt{2}\mathrm{erf}^{-1}(P(\log R < \log R^\prime)),
\end{aligned}
\end{equation}
where a value of $C = 3$ indicates a 3$\sigma$ agreement between the datasets. If $C$ becomes very large, then we would conclude that the data sets are in a suspicously high degree of agreement (see \cref{fig:interpretation}).

\section{Validating the NRE}
\label{sec:validation}

To demonstrate the robustness of our method, and some of its limitations, we first look at an example with an analytically tractable distribution of in concordance $\log R$ and compare this with the prediction from the NRE.

We begin by defining our prior and likelihood function in our example to be Gaussian and use a linear model for each of our observed datasets,
\begin{equation}
\begin{aligned}
    D = & M \theta + m \pm \sqrt{C} \\
    \mathcal{L}(D|\theta) = &\mathcal{N}(M \theta + m, C) \\
    \pi(\theta) = & \mathcal{N}(\mu, \Sigma)
\end{aligned}
\end{equation}
where $\theta$ are the model parameters, $M$ and $m$ define the data model and data samples can be drawn from the likelihood with covariance $C$. In the example that follows $M$, $m$ and $C$ are different for each experiment. $\mu$ and $\Sigma$ are the mean and covariance of our prior. In such a set-up the Bayesian evidence for each experiment and the joint observation is analytically tractable. For each experiment, the evidence is given by
\begin{equation}
    \mathcal{Z} = \mathcal{N}(m + M\mu, C + M\Sigma M^T).
\end{equation}
where $M^T$ is the transpose of $M$. We use the \textsc{lsbi} package to evaluate these expressions \footnote{\url{https://github.com/handley-lab/lsbi}}.

We draw training data from $\mathcal{Z}_{AB} = P(D_A, D_B)$ for our NRE and for each pair of $D_A$ and $D_B$ in the test data we analytically calculate $R$ to build the `true' distribution that we are trying to predict with the trained NRE.

\subsection{Assessing the performance of the NRE}

The performance of NREs is known to degrade as the absolute value of the log ratio they are predicting increases. Therefore, we might expect the performance of the \textsc{tensionnet} to degrade with increasing prior width, and we test this by comparing the prediction from the NRE with the true distribution for a range of prior widths $\Sigma$.

We define $M$ to be a matrix of uniform random numbers between 0 and 1 of dimensions $d \times n$ where $d=50$ is the number of data points and $n=3$ is the number of dimensions. $m$ and $\mu$ are defined to be a vector of uniform random numbers between 0 and 1 of length $d$ and $C$ is a diagonal matrix of $0.01$. Where $M$ and $m$ vary, the prior defined by $\Sigma$ and $\mu$ is the same for both experiments. $\Sigma$ is a diagonal matrix, and we consider three different scenarios where $\Sigma = 0.1 \mathcal{I}, 1\mathcal{I}$ and $100\mathcal{I}$ where $\mathcal{I}$ is the identity matrix. 

For each $\Sigma$ we generate $500,000$ matched observations from experiment $A$ and experiment $B$ for training the NRE. We use an exponentially decaying learning rate with an initial value of $10^{-3}$, a step size of 1000 epochs and a decay rate of 0.9. We use a ReLU activation function in the hidden layers, five hidden layers of 25 nodes each, a maximum number of epochs of 1000 with early stopping and a batch size of 1000. We use the ADAM optimizer for training. Once trained, we generate a new set of 5000 previously unseen in concordance observations from the models to put through the NRE and generate a predicted distribution of $\log R$.

\begin{figure*}
    \centering
    \includegraphics{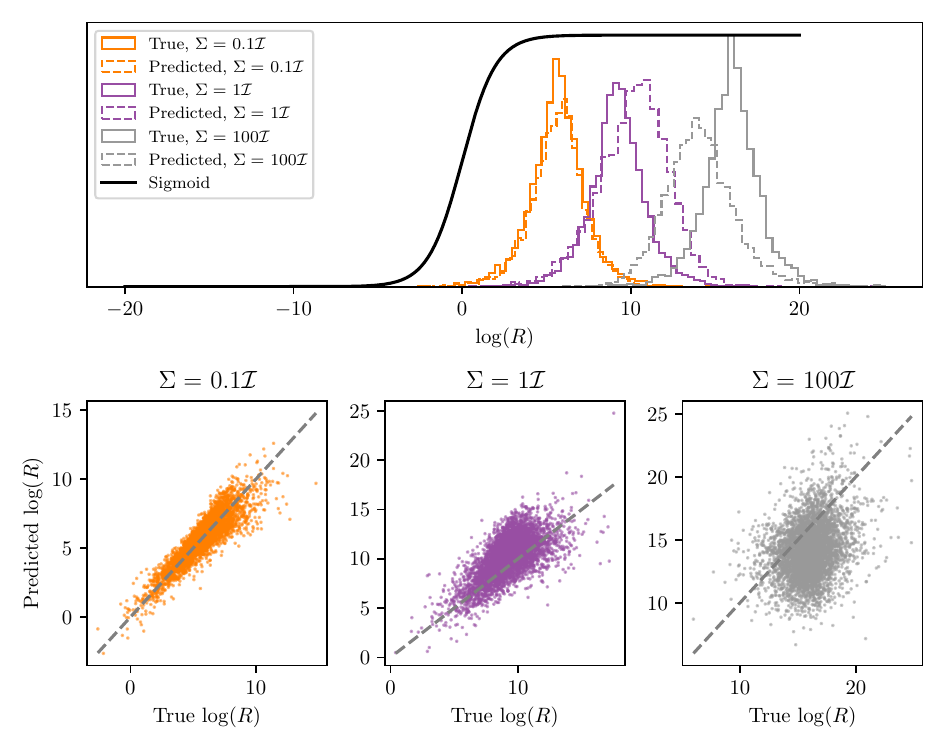}
    \caption{We hypothesise two experiments observing data that can be described with a linear model and a Gaussian likelihood function. By then defining our prior to also be Gaussian with a diagonal covariance $\Sigma$ we can analytically calculate the joint and individual evidences and the tension statistic $R$. We draw a test set from the joint distribution $\mathcal{Z}(D_A,D_B)$ which we use to analytically derive the in concordance $\log R$ distribution (solid lines, top panel) and predict the distribution from the NRE (dashed lines, top panel) for different prior widths. We also show the sigmoid activation function for reference. The bottom row shows the predicted versus true $\log R$ values for the test set for different prior widths. Performance begins to break down for $\log R > 10$.}
    \label{fig:validation_example}
\end{figure*}

The top panel of \cref{fig:validation_example} shows the predicted distributions (dashed lines) from the NRE versus the analytic distributions (solid lines) for different $\Sigma$. The solid black line shows the sigmoid activation function. The bottom three panels show the predicted versus true $\log R$ for each pair of data samples in the distribution. As the prior widens and $\log R$ becomes larger, the accuracy with which the distribution is recovered degrades as expected. Performance drops off, particularly for large prior widths, when $\log R > 10$. Caution needs to therefore be taken when using the NRE to calibrate values of $\log R \gg 10$. In such circumstances, one could consider reducing the width of the prior or running nested sampling on a handful of simulations to gauge how well the NRE is performing. If all one is interested in is the tension between different data sets, one could also choose one's prior so that $\log R$ is closer to 1 since the proposed tension metrics $T$ and $C$ are prior independent. For the orange distribution with $\Sigma=0.1\mathcal{I}$ and to some extent the purple distribution with $\Sigma = 1\mathcal{I}$, the NRE accurately recovers the $\log R$ distribution.

\subsection{Calibrating out the prior}

\begin{figure*}
    \centering
    \includegraphics[width=0.9\linewidth]{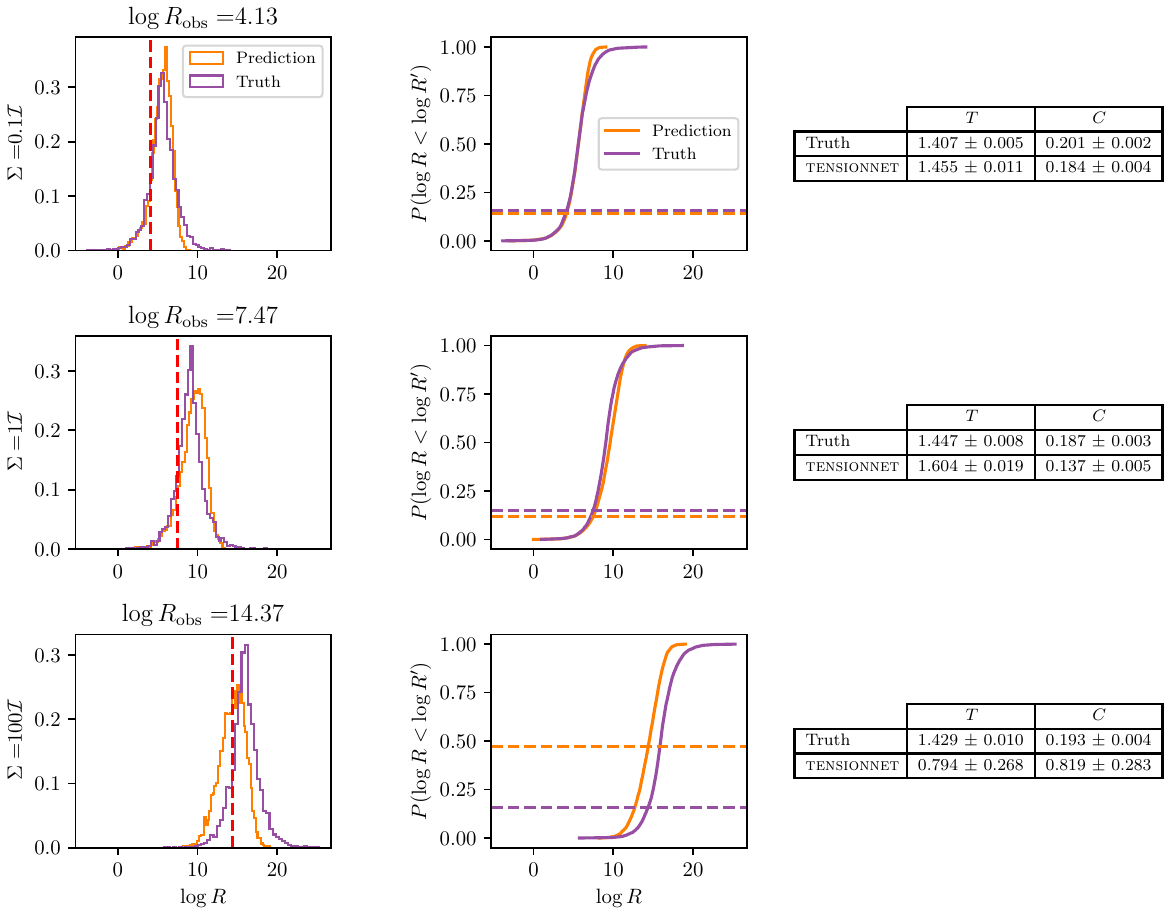}
    \caption{Using the linear model described in \cref{sec:validation} we show how the in concordance $\log R$ distribution can be used to calibrate the prior dependence of the $R$ statistic. We also show how the predicted in concordance $R$ distribution from the \textsc{tensionnet} is largely consistent with the analytic distribution. The narrowest prior is on the top row and the widest on the bottom row. The first column shows the distribution of in concordance $\log R$ values calculated analytically in purple and as predicted by the NRE in orange. We also show the analytically calculated value of $\log R$ for a simulation drawn from the narrow prior as a red dashed line. The middle column shows the CDFs derived from the two in concordance distributions and as horizontal dashed lines the value of the CDF at $R_\mathrm{obs}$ according to the analytic (purple) and NRE (orange) distributions. The final column shows the average values over five runs of $T$ and $C$ derived using the true analytic distributions and the NRE for each prior with an associated error.}
    \label{fig:lsbi-pior-dependence}
\end{figure*}

Using the above example, we can also illustrate how the \textsc{tensionnet} can be used to calibrate out the dependence of $R$ on the prior. In \cref{fig:lsbi-pior-dependence}, we keep our data model the same but change the prior width on our three parameters. Our observed dataset is drawn from the narrowest prior and kept the same throughout. We can clearly see that as the prior width increases, so does $R_\mathrm{obs}$ as expected. However, we can also see that the true distribution (purple) of in concordance $\log R$ values also shifts to higher values. When we use this distribution to calibrate $\log R_\mathrm{obs}$ into $T$ and $C$ the values are approximately constant regardless of the prior width. Calibrating against the predicted distribution from the NRE (orange) gives largely consistent results with some degradation in performance for the largest prior width as expected from the last section. We repeat the analysis five times and report the average values of $T$ and $C$ with an associated error for both the true and predicted distributions in \cref{fig:lsbi-pior-dependence}.

\section{Cosmological Examples}
\label{sec:cosmological_examples}
\subsection{Toy 21-cm Cosmology}
\label{sec:toy_21cm}

Observers in the field of 21-cm Cosmology are aiming to detect an information rich redshifted signal from neutral hydrogen from the Cosmic Dawn and Epoch of Reionization \citep[see][for reviews of the field]{Furlanetto2006, Barkana2016, Mesinger2019, Liu2020}. The signal is observed in the radio band, and  can in theory be detected as a sky-averaged 21-cm signal \citep[e.g.][]{EDGES, SARAS3, Acedo2022REACH}. It has a complex dependence on the astrophysics of the early Universe \citep[e.g.][]{Mirocha2014ARES, Mesinger201121cmfast, Reis2020radioGalaxies, Reis2021Lyalpha, GesseyJones2022IMF, Skider2024LineSight, Pochinda2023Joint, GesseyJones2024Joint, Munoz2023Zeus, Mittal2025Echo21} and can be thought of as a spectral distortion in the CMB. The key challenge in 21-cm cosmology is the separation of this signal from the dominant Galactic and extragalactic foregrounds, that the instruments also observe, whilst accounting for the non-uniform response of the instruments to the sky \citep{Anstey2021REACH}.

Since we are focused on illustrating the performance of the \textsc{tensionnet}, we ignore the effects of foregrounds and the instrument in our example and use a simplistic gaussian absorption feature to model the 21-cm signal.  We include Gaussian distributed noise in our simulated data (inspired by current observations \cite{EDGES, SARAS3}) and a Gaussian absorption feature
\begin{equation}
    \delta T_b = -A \exp \bigg( -\frac{(\nu - \nu_0)^2}{w^2} \bigg),
    \label{eq:gaussian_signal}
\end{equation}
where $A$ corresponds to the amplitude of the signal, $\nu_0$ to the central frequency and $w$ to the width. A number of works in the literature have adopted this model for illustrative purposes \citep[e.g][]{Anstey2021REACH, Bevins2021maxsmooth} but we stress that it is simplistic and does not capture the complex astrophysics encoded by the 21-cm signal.

Current observations of the sky-averaged 21-cm signal include a tentative detection by the EDGES collaboration \cite{EDGES} and an upper limit on the magnitude of the signal from SARAS3~\cite{SARAS3}. Analysis by the SARAS3 team suggested that these measurements are in tension with each other, and a number of works have discussed the possible presence of systematics in the EDGES data \cite{Hills2018Edges, Singh2019EDGES, Sims2020EDGES, Bevins2021maxsmooth}. As more experiments come online in the coming years \citep[e.g.][]{Acedo2022REACH} the assessment of tension and concordance between different observations is going to become crucial for the field.

\comment{
\begin{table}[]
    \centering
    \begin{tabular}{|c|c|c|}
    \hline
         $A_B/A_A$ & $T$ & $C$\\
         \hline
         \hline
         0.75 & $2.154^{+0.073}_{-0.063}$ & $0.039^{+0.008}_{-0.006}$ \\
         \hline
         1.00 & $0.043^{+0.018}_{-0.027}$ & 
         $2.120^{+0.218}_{-0.207}$\\
         \hline
         1.25 &$3.35^{+0.158}_{-0.001}$ & $0.001 \pm 0.001$\\
         \hline
    \end{tabular}
    \caption{Caption}
   \label{tab:my_label}
\end{table}
}

\begin{figure*}
    \centering
    \includegraphics[width=0.9\linewidth]{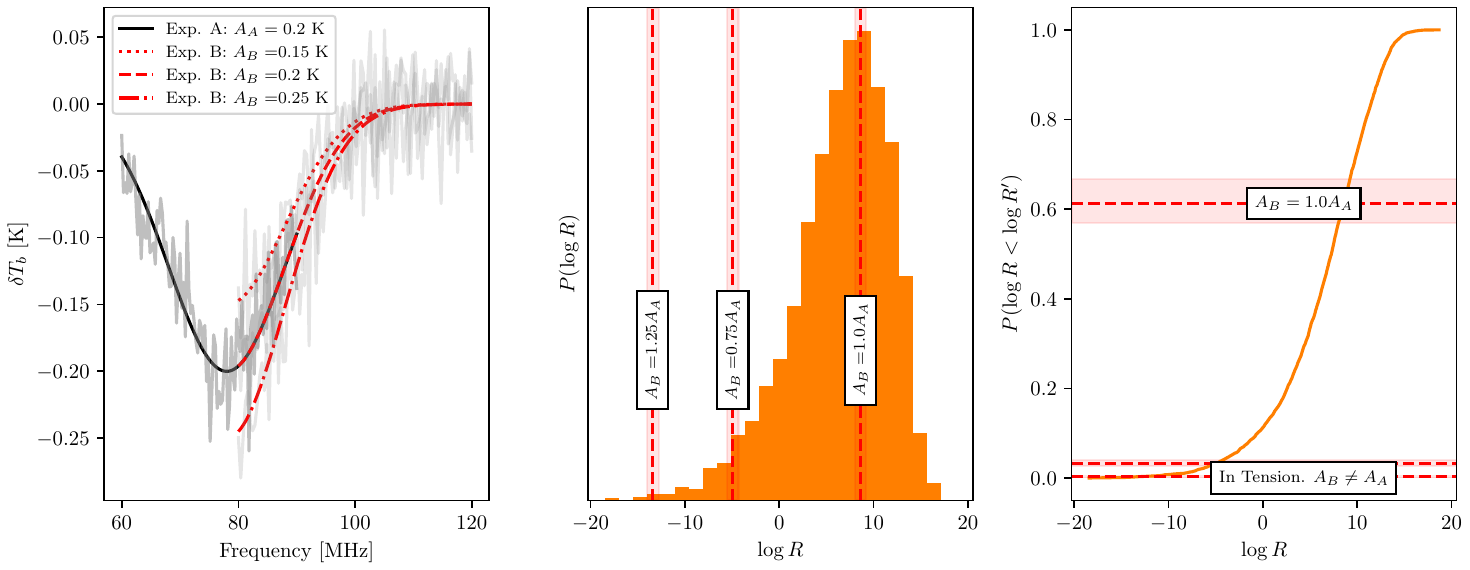}
    \caption{To further illustrate the application of NREs to the calibration of $R$ we use a toy example inspired by 21-cm cosmology. \textbf{Left Panel:} We simulate an experiment observing a Gaussian absorption trough as a function of frequency (black line) and three different scenarios in which another experiment measures a 21-cm signal with either the same or different amplitudes in a different band (red lines). To each observation, we add Gaussian random noise with a standard deviation of 25 mK (shown in grey and motivated by current observations \citep[e.g.][]{EDGES}). \textbf{Middle Panel:} We train the NRE on simulated observations of the signal by both experiments, covering a wide prior range of signal parameters. We use the NRE to evaluate the possible distribution of in concordance $\log R$ values. We plot the observed $\log R$ for each pair of observations from experiment A and B. \textbf{Right Panel:} Finally, we show the CDF of the in concordance $\log R$ distribution in the right panel of the figure and the corresponding CDF values for each pair of observations. We find that for the two in tension observations the  $T=2.989^{+0.167}_{-0.060}$ and $T = 2.147^{+0.056}_{-0.089}$ and for the in concordance observations $C = 0.864^{+0.107}_{-0.076}$ and $T = 0.507^{+0.063}_{-0.078}$.}
    \label{fig:toy_example}
\end{figure*}

We simulate observations of the sky-averaged 21-cm signal from two different experiments in different frequency ranges. We hypothesise that the 21-cm signal has a depth of 0.2 K, a central frequency of 78 MHz and a width of 10 MHz. In our example, the first experiment (Exp. A) has made a detection of the signal with Gaussian distributed noise with a standard deviation of 25~mK over the frequency range $60-90$ MHz with a channel width of $\approx 0.3$~MHz (see top left panel of \cref{fig:toy_example}). We then hypothesise a series of scenarios where a second experiment (Exp. B) has observed the 21-cm signal in the frequency range $80 - 120$ MHz with a channel width of $\approx 0.4$~MHz with the same central frequency and width but a different magnitude $A = [ 0.15, 0.2, 0.25]$ K such that the observations are in tension, concordance and tension respectively. We add 25 mK Gaussian random noise to the data from experiment B.
 
We fit each pair of observations using the nested sampling implementation \textsc{polychord} \citep{Handley2015polychorda, Handley2015polychordb} to assess $R_\mathrm{obs}$. We use \cref{eq:gaussian_signal} as our model, $M$ for the data, $D$ and use a Gaussian likelihood function
\begin{equation}
    \log \mathcal{L} = \sum_i -\frac{1}{2} \log 2\pi \sigma^2 - \frac{1}{2} \frac{(D_i - M_i)^2}{\sigma^2},
\end{equation}
where the sum is over observation frequency and $\sigma$ is the standard deviation of the noise, which we fit as a free parameter. The prior is uniform on $A$ between $0.0 - 4.0$ K, $\nu_0$ between $60 -80$ MHz, $w$ between $5 - 40$ MHz and $\sigma$ between $0.001 - 0.1$~K. 

The combination of our likelihood and prior and the fact that our model is non-linear makes the in concordance $\log R$ distribution analytically intractable. It can only be accessed in a reasonable amount of time through the \textsc{tensionnet}. One could of course evaluate the distribution with 1000s of Nested Sampling runs, but this would be computationally expensive. To calculate $R$ using Nested Sampling for this example costs around 12 seconds (all the analysis performed in the paper was done on an M2 MacBook with 8GB ram). This includes the run time of the inference on each data set individually and on the joint data set. Using Nested Sampling to evaluate $R$ for 5000 pairs of simulated in concordance, observations to assess the distribution of possible values, would therefore take around 16.7 hrs. To generate simulations for the NRE, train the NRE and evaluate it for 5000 pairs of data sets to get the in concordance $R$ distribution takes around 6.6 minutes. For this problem, the NRE is therefore $(16.7 \times 60)/6.6 \approx 152$ times faster. Note that this is a trivial problem with a fast likelihood function.

We generate 200,000 mock observations of the 21-cm signal for both experiments with the same sets of parameters from the prior. We then shuffle these datasets to create a corresponding set of in tension `observations' giving us a set of 400,000 simulations. We use $80\%$ of this to train the NRE and the rest to perform early stopping.

Once trained, we generate 5000 pairs of observations of the same signal by both experiment with parameters drawn randomly from the prior range to evaluate the in concordance  $\log R$ distribution. From this distribution, we can calculate an empirical CDF and compare the observed $R$ statistic for the three pairs of observations. Nested sampling returns an error on the Bayesian evidence, which can then be propagated forward through to $\log R_\mathrm{obs}$ and the tension statistics $T$ and $C$. For the two in tension datasets we find $T=2.989^{+0.167}_{-0.060}$ and $T = 2.147^{+0.056}_{-0.089}$ and for the in concordance case when both experiments observe the same signal $C = 0.864^{+0.107}_{-0.076}$ and $T = 0.507^{+0.063}_{-0.078}$. This is in agreement with our expectations, given the amplitude of the signals in the different data sets, and demonstrates that the \textsc{tensionnet} performs well. The results are summarised in \cref{fig:toy_example}.

\subsection{DESI and SDSS}
\label{sec:toy_cosmo}

We next investigate the tension between the Baryon Acoustic Oscillations~(BAO) cosmological constraints from the Sloan Digital Sky Survey~(SDSS) \cite{Alam2015SDSSDR12, Ahumada2020SDSSDR16} and the recent Dark Energy Spectroscopic Instrument~(DESI) data release \cite{adame2024desiCosmologyBAO}.

Before recombination when photons and baryons were coupled via Thomson scattering, oscillations were set up in the hot plasma by the competing forces of gravity and radiation pressure. Spherical density perturbations in the coupled plasma propagated outwards as acoustic waves. Once the photons and the baryons decouple at recombination, these acoustic waves stop travelling through the baryon fluid and the scale of the wave is imprinted in the matter distribution. The scale of the acoustic waves at recombination is known as the sound horizon. The photons free stream and form the CMB. Since the baryons and dark matter are coupled by gravity, the acoustic waves imprint a preferential scale for structure formation and the distance between two galaxies in the later Universe. The BAO scale is hence a standard ruler, and observations of it can be used to constrain the expansion rate of and the matter density of the Universe \cite{Bassett2010BAO, Cuceu2019BAO}.

In practice, the BAO scale is estimated via the cross-correlation of the position of galaxies $\xi$ in large surveys like SDSS and DESI and shows up as a bump in $\xi(\theta)$ and $\xi(\Delta z)$ where $\theta$ is the angular separation of galaxies and $\Delta z$ the redshift separation. Angular scales on the sky $\theta$ are related to commoving physical sizes $\lambda$ by
\begin{equation}
    \theta = \frac{\lambda}{(1+z) D_A} = \frac{\lambda}{D_M},
\end{equation}
where $D_A$ is the angular diameter distance and $D_M$ is the comoving angular diameter distance also known as the transverse comoving distance. Similarly, physical size is related to redshift separation by
\begin{equation}
    \Delta z = \frac{\lambda H(z)}{c} = \frac{\lambda}{D_H},
\end{equation}
where $D_H$ is the Hubble distance, $c$ is the speed of light and $H(z)$ is the Hubble constant as a function of redshift. From $\xi(\theta)$ and $\xi(\Delta z)$ we can approximate the angular size of the BAO at a given redshift $\theta_\mathrm{BAO}$ and, given a large enough set of galaxy measurements as a function of redshift, the redshift separation $\Delta z_\mathrm{BAO}$. The comoving size of the BAO is equal to the sound horizon $\lambda = r_s$ at recombination when the photons and baryons decouple. BAO observations therefore give us a measure of $D_M/r_s$ and $D_H/r_s$ from which we can constrain cosmology.

The BAO signature appears in the cross-correlation of a number of different objects such as Luminous Red Galaxies~(LRG), Emission Line Galaxies~(ELG), quasars and the Lyman-$\alpha$ forest. Each class of objects probes a different redshift range, and measurements of $D_M/r_s$ and $D_H/r_s$ are ascribed to an effective redshift \cite{Bassett2010BAO}.

We generate theoretical models for the observables with \textsc{CAMB} \citep{Lewis1999CAMB, Lewis2002CAMB}, then taking advantage of the reported covariance estimates for the SDSS and DESI observations use analytic likelihoods, implemented with \textsc{scipy}, to generate noisy observations of the theory model.

There is a partial overlap in the redshift range and sky coverage of SDSS and DESI, and as such the full datasets include some of the same galaxies. Therefore, the surveys are correlated and if we want to perform a joint Bayesian inference of the datasets with tools like nested sampling to recover $R_\mathrm{obs}$ we need a joint likelihood function. Although the level of correlation between the datasets has been estimated \cite{adame2024desiLyaBAO, adame2024desiCosmologyBAO}, the derivation of a joint likelihood is beyond the scope of this paper and has not yet been attempted in the literature.

\begin{figure*}
    \centering
    \includegraphics[width=0.9\linewidth]{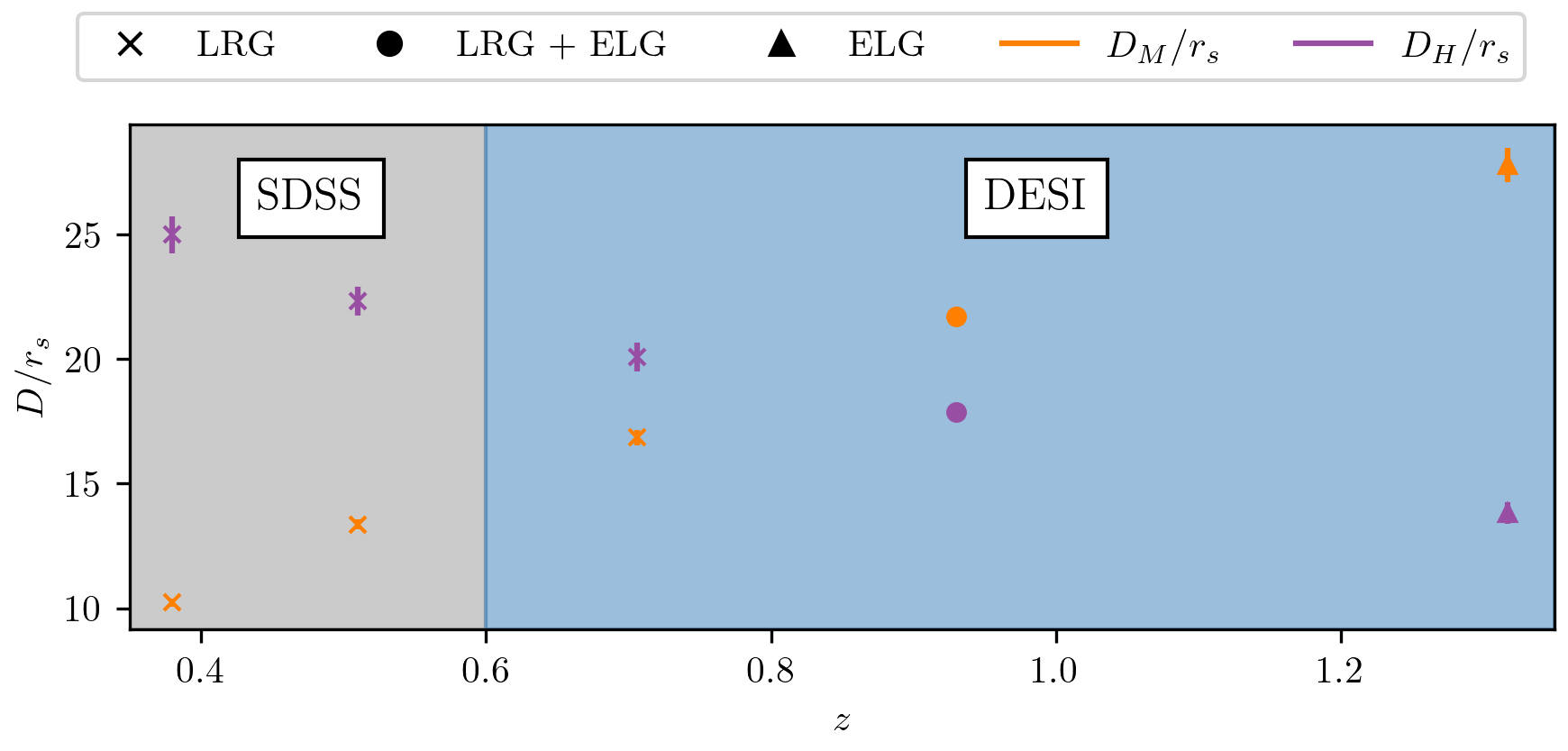}
    \caption{The composite BAO dataset used in this work from SDSS and DESI observations. Following the discussion in \cite{adame2024desiCosmologyBAO} we curate the dataset by taking SDSS observations of the BAO scale from Luminous Red Galaxies~(LRG) below $z=0.6$ (grey shaded region) and high redshift observations of Luminous Red Galaxies and Emission Line Galaxies~(ELG) from DESI (blue shaded region).}
    \label{fig:BAO-data}
\end{figure*}

An alternative approach is to build a joint SDSS and DESI dataset by selecting data points from one survey or the other at each sampled effective redshift. In \cite{adame2024desiCosmologyBAO} the authors demonstrate this idea by selecting SDSS observations below $z=0.6$ and DESI observations above $z=0.6$ to maximise the effective volume covered by the joint dataset. In our analysis we use
\begin{itemize}
    \item SDSS LRG at $z_\mathrm{eff} = 0.38$ and $0.51$
    \item DESI LRG at $z_\mathrm{eff} = 0.706$
    \item DESI LRG-ELG at $z_\mathrm{eff} = 0.930$
    \item DESI ELG at $z_\mathrm{eff} = 1.317$
\end{itemize}
and the combined dataset is shown in \cref{fig:BAO-data}.
A more complete analysis can be pursued in the future when correlated likelihoods become available.
Some tension, at approximately 3$\sigma$ level, has been observed between SDSS and DESI at an effective redshift of $z_\mathrm{eff}\approx0.7$ \cite{adame2024desiTension},
although this was not arrived at via a joint analysis but rather an assessment of the individual measurements and the correlation between the datasets. We do not expect to see this tension in our analysis, as we are just considering the DESI measurement at $z_\mathrm{eff}\approx0.7$.

We constrain the baryon density $\Omega_b h^2$, dark matter density $\Omega_c h^2$, the slope and amplitude of the matter power spectrum $n_s$ and $\log 10^{10} A_s$ and the value of $h = \frac{H_0}{100~{\rm km~s}^{-1}~{\rm Mpc}^{-1}}$. We fix the value of $\tau$ to the best fit value from the Planck 2018 analysis of $0.055$ \cite{Planck2018}. The prior is uniform on $\Omega_b h^2$ between $0.01 - 0.085$, $\Omega_c h^2$ between $0.08 - 0.21$, $n_s$ between $0.8 - 1.2$, $\log 10^{10} A_s$ between $2.6 - 3.8$ and $h$ between $0.5 - 0.9$. It is motivated by the prior in \cite{Handley_tensions_2019}, which is somewhat motivated by the default priors for \textsc{CosmoMC}~\footnote{https://cosmologist.info/cosmomc/}, and designed to encompass the Planck and Dark Energy Survey Y1 posteriors. As discussed in \cite{Handley_tensions_2019}, however, there is nothing particularly special about this prior and in practice it could be broadened or narrowed without causing any objections in the community.
For each measurement of the BAO signature $D$ our likelihood is Gaussian, as in \cite{Cuceu2019BAO}, with a covariance given by the measured covariance $\Sigma$.

The SDSS data is available at \url{https://www.sdss4.org/dr17/} and the DESI data has been reported in \cite{adame2024desiCosmologyBAO}. Both datasets have been collected together as part of the \textsc{COBAYA} cosmological likelihood code \footnote{\url{https://cobaya.readthedocs.io/en/latest/likelihood_bao.html}}. Using nested sampling and \textsc{CAMB}, we find $\log R_\mathrm{obs} =2.57 \pm 0.30$. Since $R_{\rm obs} < 10$ we are not worried about the NRE saturation that was previously discussed.

To train the NRE, we generate 100,000 examples of in concordance observations from SDSS and DESI. We then separate out 10\% of these for testing and shuffle the remaining $90\%$ to create a set of 180,000 matched and mismatched observations. These are then split into training and validation datasets of 120,600 and 59,400 (33\%) observations respectively. We use an exponentially decaying learning rate scheduler with an initial learning rate of $10^{-3}$, a step size of $1000$ epochs and a decay rate of $0.9$. We train for a maximum of 1000 epochs with a batch size of 1000 and a patience of 50. We use L1 kernel regularization to improve the performance. We standardize the simulations at each redshift using the mean and standard deviation of the training data.

\begin{figure*}
    \centering
    \begin{tikzpicture}[rednode/.style={circle, draw=red!60, fill=red!5, very thick,                 minimum size=3mm, node distance=0.1cm and 1cm},
                bluenode/.style={circle, draw=blue!60, fill=blue!5, very thick, minimum size=3mm, node distance=0.1cm and 1cm},
                greennode/.style={circle, draw=green!60, fill=green!5, very thick, minimum size=3mm},
                remember picture]
        \node (0) {};

        \node[rednode, above=of 0, align=center](layer1_center1) {};
        \node[rednode, below=of layer1_center1, align=center](layer1_center2) {};
        \node[rednode, above=of layer1_center1, align=center](layer1_top2) {};
        \node[rednode, below=of layer1_center2, align=center](layer1_bottom1) {};

        \node[rednode, left=of layer1_top2, align=center, ](hl1) {};
        \node[rednode, left=of layer1_center1, align=center](hl2) {};
        \node[rednode, left=of layer1_center2, align=center](hl3) {};
        \node[rednode, left=of layer1_bottom1, align=center](hl4) {};
        
        \node[bluenode, left=of hl1, align=center, yshift=0.25cm](compress_out1) {};
        \node[bluenode, above=of compress_out1, align=center](compress_out2) {};

        \node[bluenode, above left=of compress_out1](compress1_layer23){};
        \node[bluenode, above =of compress1_layer23](compress1_layer22){};
        \node[bluenode, above =of compress1_layer22](compress1_layer21){};
        \node[bluenode, below =of compress1_layer23](compress1_layer24){};
        \node[bluenode, below =of compress1_layer24](compress1_layer25){};

        \node[bluenode, left=of compress1_layer23](compress1_layer13){};
        \node[bluenode, left=of compress1_layer22](compress1_layer12){};
        \node[bluenode, left=of compress1_layer21](compress1_layer11){};
        \node[bluenode, left=of compress1_layer24](compress1_layer14){};
        \node[bluenode, left=of compress1_layer25](compress1_layer15){};

        \node[bluenode, left=of compress1_layer13](compress1_in3){};
        \node[bluenode, left=of compress1_layer12](compress1_in2){};
        \node[bluenode, left=of compress1_layer11](compress1_in1){};
        \node[bluenode, left=of compress1_layer14](compress1_in4){};
        \node[bluenode, left=of compress1_layer15](compress1_in5){};

        \foreach \x in {compress1_layer21, compress1_layer22, compress1_layer23, compress1_layer24, compress1_layer25}{
        \foreach \y in {compress_out1, compress_out2}{
        \draw[->](\x.east)--(\y.west);}}

        \foreach \x in {compress1_layer11, compress1_layer12, compress1_layer13, compress1_layer14, compress1_layer15}{
        \foreach \y in {compress1_layer21, compress1_layer22, compress1_layer23, compress1_layer24, compress1_layer25}{
        \draw[->](\x.east) -- (\y.west);}}

        \foreach \x in {compress1_in1, compress1_in2, compress1_in3, compress1_in4, compress1_in5}{
        \foreach \y in {compress1_layer11, compress1_layer12, compress1_layer13, compress1_layer14, compress1_layer15}{
        \draw[->](\x.east) -- (\y.west);}}
        
        \node[bluenode, left=of hl4, align=center, yshift=-0.25cm](compress2_out1) {};
        \node[bluenode, below=of compress2_out1,  align=center](compress2_out2) {};

        \node[bluenode, below left=of compress2_out1](compress2_layer23){};
        \node[bluenode, above =of compress2_layer23](compress2_layer22){};
        \node[bluenode, above =of compress2_layer22](compress2_layer21){};
        \node[bluenode, below =of compress2_layer23](compress2_layer24){};
        \node[bluenode, below =of compress2_layer24](compress2_layer25){};

        \node[bluenode, left=of compress2_layer23](compress2_layer13){};
        \node[bluenode, left=of compress2_layer22](compress2_layer12){};
        \node[bluenode, left=of compress2_layer21](compress2_layer11){};
        \node[bluenode, left=of compress2_layer24](compress2_layer14){};
        \node[bluenode, left=of compress2_layer25](compress2_layer15){};

        \node[bluenode, left=of compress2_layer13](compress2_in3){};
        \node[bluenode, left=of compress2_layer12](compress2_in2){};
        \node[bluenode, left=of compress2_layer11](compress2_in1){};
        \node[bluenode, left=of compress2_layer14](compress2_in4){};
        \node[bluenode, left=of compress2_layer15](compress2_in5){};

        \foreach \x in {compress2_layer21, compress2_layer22, compress2_layer23, compress2_layer24, compress2_layer25}{
        \foreach \y in {compress2_out1, compress2_out2}{
        \draw[->](\x.east)--(\y.west);}}

        \foreach \x in {compress2_layer11, compress2_layer12, compress2_layer13, compress2_layer14, compress2_layer15}{
        \foreach \y in {compress2_layer21, compress2_layer22, compress2_layer23, compress2_layer24, compress2_layer25}{
        \draw[->](\x.east) -- (\y.west);}}

        \foreach \x in {compress2_in1, compress2_in2, compress2_in3, compress2_in4, compress2_in5}{
        \foreach \y in {compress2_layer11, compress2_layer12, compress2_layer13, compress2_layer14, compress2_layer15}{
        \draw[->](\x.east) -- (\y.west);}}
        
        \node[greennode, right=of layer1_center1, text width=0.7cm, align=center, yshift=-0.25cm](output_1) {$\log R$};

        \foreach \x in {compress_out1, compress_out2, compress2_out1, compress2_out2}{
            \foreach \y in {hl1, hl2, hl3, hl4}{
            \draw[->](\x.east) -- (\y.west);}}
        
        \foreach \x in {hl1, hl2, hl3, hl4}{
            \foreach \y in {layer1_center1, layer1_center2, layer1_top2, layer1_bottom1}{
                \draw[->](\x.east) -- (\y.west);}}
        
        \draw[->](layer1_center1.east) -- (output_1.west);
        \draw[->](layer1_center2.east) -- (output_1.west);
        \draw[->](layer1_top2.east) -- (output_1.west);
        \draw[->](layer1_bottom1) -- (output_1.west);

        \foreach \x/\label/\c in {compress1_in1/\tiny SDSS LRG $z_\mathrm{eff}=0.38$/red!20, compress1_in2/\tiny SDSS LRG $z_\mathrm{eff}=0.51$/red!20, compress1_in3/\tiny DESI LRG $z_\mathrm{eff}=0.706$/blue!20, compress1_in4/\tiny DESI LRG-ELG $z_\mathrm{eff}=0.930$/blue!20, compress1_in5/\tiny DESI LRG $z_\mathrm{eff}=1.317$/blue!20}{
        \node[draw, rectangle, left=of \x, text width=3cm, align=center, node distance=0.1cm and 0.1cm, fill=\c, xshift=0.9cm]{\label};
        }

        \foreach \x/\label/\c in {compress2_in1/\tiny SDSS LRG $z_\mathrm{eff}=0.38$/red!20, compress2_in2/\tiny SDSS LRG $z_\mathrm{eff}=0.51$/red!20, compress2_in3/\tiny DESI LRG $z_\mathrm{eff}=0.706$/blue!20, compress2_in4/\tiny DESI LRG-ELG $z_\mathrm{eff}=0.930$/blue!20, compress2_in5/\tiny DESI LRG $z_\mathrm{eff}=1.317$/blue!20}{
        \node[draw, rectangle, left=of \x, text width=3cm, align=center, node distance=0.1cm and 0.1cm, fill=\c, xshift=0.9cm]{\label};
        }

        \node[rectangle, left=of compress1_in3, text width=1cm, align=center, node distance=0.1cm and 0.1cm, xshift=-3cm](dm){$D_M/r_s$};

        \node[ rectangle, right=of dm, text width=1cm, align=center, node distance=0.1cm and 0.1cm, xshift=-1.4cm]{\Huge \{};

        \node[rectangle, left=of compress2_in3, text width=1cm, align=center, node distance=0.1cm and 0.1cm, xshift=-3cm](dh){$D_H/r_s$};

        \node[rectangle, right=of dh, text width=1cm, align=center, node distance=0.1cm and 0.1cm, xshift=-1.4cm]{\Huge \{};

    \end{tikzpicture}
    \caption{An exact diagram of hidden layer structure in the DESI-SDSS \textsc{tensionnet}. We find that combining and compressing the information in the measurements of $D_M/r_s$ and $D_H/r_s$ from DESI and SDSS into a latent space before mixing information from the two measurements improves the performance of the NRE. The compression networks (in blue) and the NRE (in red) are trained together under the same binary cross entropy loss function.}
    \label{fig:bao-nre}
\end{figure*}
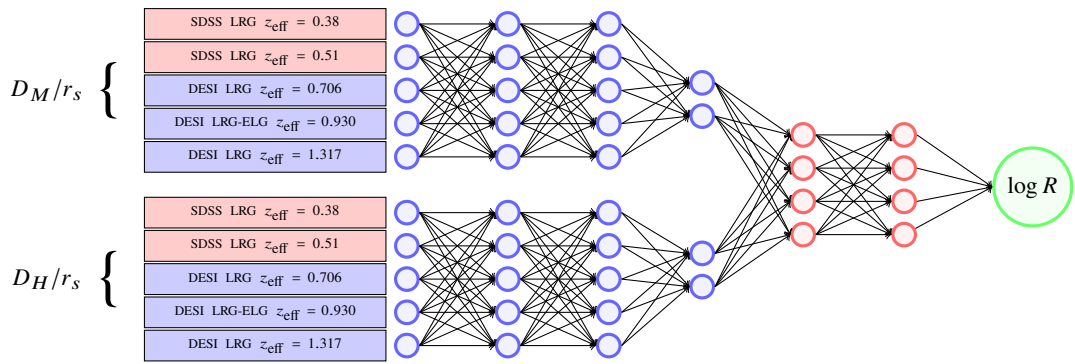

We group together the measurements of $D_M/r_s$ from DESI and SDSS at the different effective redshifts and compress them down into a smaller latent space. We do the same with the measurements of $D_H/r_s$ before passing them to the NRE. We find that compressing the data in this way works better than directly passing the raw data to the NRE. This initial step of keeping the measurements of $D_M/r_s$ and $D_H/r_s$ separate allows the NRE to learn the trends, like those seen in \cref{fig:BAO-data}, in each variable as a function of redshift before mixing information from the two together. We found that smaller neural network architectures worked better for this problem, as there are only 10 data points in the combined set. The compression networks have three layers of 5, 5 and 2 hidden nodes and the NRE has 2 layers of 4 nodes each. The compression layers and the NRE are trained together. The architecture of the network can be seen in \cref{fig:bao-nre}.

\begin{figure}
    \centering
    \includegraphics{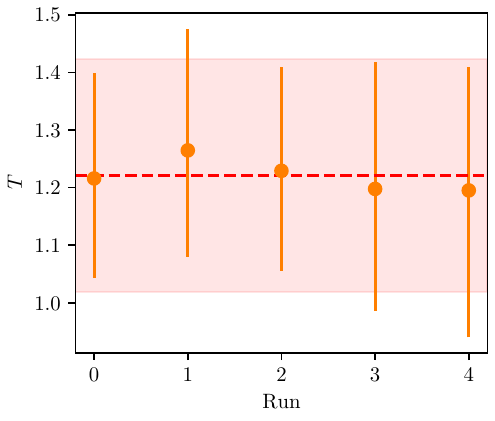}
    \caption{We repeat training of our NRE five times on simulations of the data from SDSS and DESI and use the predicted distributions to evaluate $T$. If our network was poorly converged, or we had too little training data, then the recovered distribution of in concordance $R$ values would vary significantly. As a result, the calculated value of $T$ would be inconsistent, and we would see a large scatter in the reported values. Instead, we see that, for the SDSS+DESI analysis, $T$ is consistent across the different training runs. On average, we find that $T = 1.22 \pm 0.20$.}
    \label{fig:bao_sigmaD}
\end{figure}

\begin{figure*}
    \centering
    \includegraphics[width=0.9\linewidth]{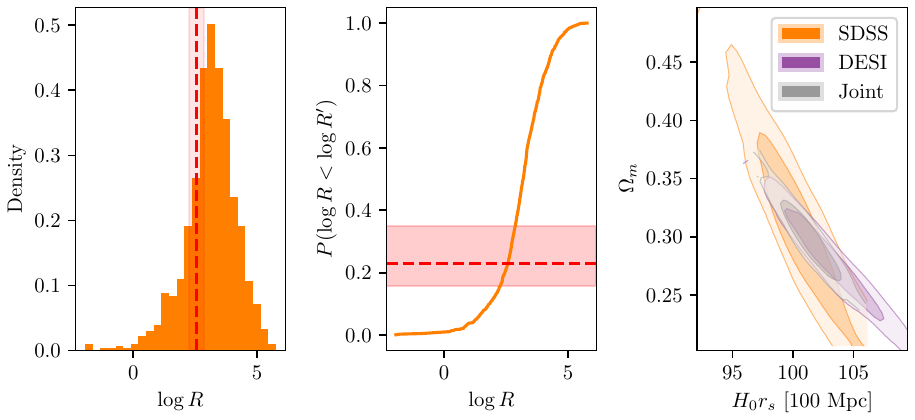}
    \caption{\textbf{Left Panel:} The predicted distribution of in concordance $\log R$ values for the curated SDSS and DESI dataset analysed in this work. The red dashed line shows the value of $\log R_\mathrm{obs}$ for the observed data calculated with nested sampling with an associated error. \textbf{Middle Panel:} The CDF corresponding to the in concordance $\log R$ distribution. Calibrating $\log R_\mathrm{obs}$ (red dashed line and shaded region) in to $\sigma$s of tension gives $T = 1.23^{+0.21}_{-0.20}$. \textbf{Right Panel:} The constraints on the matter overdensity $\Omega_m$ and the combination of the Hubble constant $H_0$ and sound horizon $r_s$ from analysis of the DESI and SDSS datasets. Our results are slightly different to those presented in Fig. 2 of \cite{adame2024desiCosmologyBAO} because we have used a different prior and not included the quasar measurements from DESI or the Lyman-$\alpha$ measurements from both surveys.}
    \label{fig:bao-calibration}
\end{figure*}

We train the network on the same training data five times with different random initial seeds to assess the consistency of our results. The corresponding values of $T$ are shown in \cref{fig:bao_sigmaD}. We find that on average, $T = 1.22 \pm 0.20$ between the combined SDSS and DESI datasets. We show an example of the calibration performed for one of the training and calibration runs in \cref{fig:bao-calibration} along with the constraints on $\Omega_m$ and $H_0 r_s$. We find no significant tension between the SDSS measurements of the BAO scale at $z_\mathrm{eff} = 0.38$ and $0.51$ and the DESI measurements at higher redshifts of $z_\mathrm{eff} = 0.706, 0.930$ and $1.317$.

\section{Limitations}
\label{sec:limitations}

As with all simulation based inference methods, the success of the \textsc{tensionnet} is dependent on how well the simulations represent the true observed datasets. In some respects, the method is also limited by the need for simulations. For example, to assess the tension between supernova observations of $H_0$ and CMB measurements using the $R$ statistic and the \textsc{tensionnet} one would need to be able to simulate the observations in a consistent framework. While work is being pursued in this direction \citep[e.g.][]{Watts2023Cosmoglobe, Karchev2024supernova} it is a notoriously difficult problem.

It is also currently difficult to verify the output of the \textsc{tensionnet}. In practice, one could run a coverage test on the recovered distribution of $\log R$ \cite{Lemos2023CoverageTest}. However, this only tells you how self-consistent the recovered distribution is and not whether it is centred around the correct $\log R$ value. One way to test this is to take a number of simulated datasets in the predicted distribution and calculate their $\log R$ value via an independent method such as nested sampling. An alternative validation approach is to repeat the NRE training to check for stability as in \cref{fig:bao_sigmaD}.

As demonstrated in \cref{sec:validation}, the \textsc{tensionnet} is limited by the NREs ability to predict extreme values of $\log R$. Sensible choices of prior distributions can help alleviate this issue, and the validation methods discussed above can help build confidence in the predicted distribution.

\section{Conclusions}
\label{sec:conclusions}

Estimating tension between different datasets is an important part of the scientific process and has become integral to the analysis of cosmological and astrophysical data. By correctly quantifying tension between different experiments, we are able to better understand our instruments and identify gaps in our knowledge. Commonly encountered examples of tension in cosmology are the $H_0$ and $\sigma_8$ tensions, although other examples exist, including in the field of 21-cm cosmology. 

A number of ways to quantify tension have been proposed including eigentension, goodness of fit degradation and Suspiciousness and these can often be translated into $\sigma$s of tension where $\sigma$ is the standard deviation of a normal distribution. A Bayesian way to quantify tension is with the tension statistic $R$ which encodes our increased confidence in one experiment's measured data given observations from another. Formerly, $R$ is the ratio of joint Bayesian evidence to the product of the individual evidences for two datasets under a common model and prior. It is symmetric, parameterisation invariant and dimensionally consistent, however, it has a non-trivial dependence on the prior. $\log R$ is typically interpreted as indicating tension if $R \ll 1$ and concordance if $R \gg 1$ or via a Jeffery's scale, neither of which properly account for the prior dependence.

For any pair of experiments observing the same physics, any model for the data and any prior distribution, there is a distribution of in concordance $\log R$ values. Having access to this distribution allows you to calibrate out the prior dependence from the observed $R$ and robustly convert the statistic into $\sigma$s of tension or concordance. Unfortunately, for most problems, this distribution is not analytically accessible. In this paper, we have shown that it can be readily accessed with simulations of the experimental observables and neural ratio estimation.

We demonstrated the application of NREs to the calibration of $R$ using toy examples and observations of the BAO scale from SDSS and DESI. By selecting observations of the BAO scale from each survey at specific effective redshifts, we avoid having to worry about the correlation between the observations whilst maximising the effective volume of the combined survey. We find no significant tension between the SDSS Luminous Red Galaxy measurements at $z_\mathrm{eff} = 0.38$ and $0.51$ and the DESI Luminous Red Galaxy measurements and Emission Line Galaxy measurements at $z_\mathrm{eff} = 0.706, 0.930$ and $1.317$. 

In \cite{adame2024desiTension} some tension has been seen between the SDSS and DESI datasets at $z_\mathrm{eff} \approx 0.7$. In practice, this could be assessed with the \textsc{tensionnet} in the future should a correlated likelihood function become available for calculating the observed $R$ with nested sampling.

Like all simulation based methods, the \textsc{tensionnet} is limited by the accuracy of the simulated observations and indeed by our ability to simulate the data in the first instance. We also find that performance of the NRE degrades as the prior widens, and sensible prior choices need to be made. We suggest that repeated training of the NRE and evaluation of $R$ for a handful of simulations with nested sampling or an independent evidence estimation tool can be done to validate the results.

We have shown that neural ratio estimators offer a cheap and effective way to access the in concordance $\log R$ distribution needed to calibrate out the prior dependence of the $R$ statistic. While acknowledging the limitations of this method, we believe it offers a promising step towards simulation based tension quantification. We expect that the method proposed in this paper will be broadly applicable beyond cosmology in other fields where tensions appear \cite[e.g.][]{CDF2022}.

\section{Acknowledgements}

HTJB acknowledges support from the Kavli Institute for Cosmology Cambridge and the Kavli Foundation. WJH thanks the Royal Society for their support through their University Research Fellowships. TGJ acknowledges the support of the Science and Technology Facilities Council (UK) through grant ST/V506606/1 and the Royal Society.

This work used the DiRAC Data Intensive service (CSD3, project number ACSP289) at the University of Cambridge, managed by the University of Cambridge University Information Services on behalf of the STFC DiRAC HPC Facility (www.dirac.ac.uk). The DiRAC component of CSD3 at Cambridge was funded by BEIS, UKRI and STFC capital funding and STFC operations grants. DiRAC is part of the UKRI Digital Research Infrastructure

\section{Data Availability}
The code and data used in this paper are available at \url{https://github.com/htjb/tension-networks}.

\bibliographystyle{apsrev4-2-titles}
\bibliography{journals, ref}

\appendix

\end{document}